\documentclass[runningheads]{apinv}
\usepackage{epsfig,cite,graphics,xcolor,ulem}
\usepackage{marvosym} 
\usepackage[utf8]{inputenc}
\usepackage{natbib,url}

\begin{document}

\title{Evolution of the optical emission lines and the X-ray emission during the super-active stage of T~CrB}
\titlerunning{Optical and X-ray emission during the super-active stage of T~CrB}
\author{K. A. Stoyanov\inst{1}, G. J. M. Luna\inst{2}, R. K. Zamanov\inst{1},
K. I{\l}kiewicz\inst{3}, \\
Y. M. Nikolov\inst{1}, M. Moyseev\inst{1}, M. Minev\inst{1}, A. Kurtenkov\inst{1,4}, S. Y. Stefanov\inst{1}}
\authorrunning{K. A. Stoyanov et al.}
\tocauthor{K. A. Stoyanov et al.} 
   \institute{Institute of Astronomy and National Astronomical Observatory, Bulgarian Academy of Sciences, 72 Tsarigradsko Chaussee Blvd., 1784 Sofia, Bulgaria
      \and
   CONICET-Universidad Nacional de Hurlingham, Av. Gdor. Vergara 2222, Villa Tesei, Buenos Aires, Argentina
   \and
   Astronomical Observatory, University of Warsaw, Al. Ujazdowskie 4, 00-478 Warszawa, Poland
   \and
   Faculty of Physics, Sofia University, 5 J. Bourchier Blvd, 1164 Sofia, Bulgaria
 }

   \date{Received September 15, 1996; accepted March 16, 1997}

\maketitle
  \abstract
{T~CrB is a symbiotic star that experiences nova outbursts every $\sim$ 80~yr. The next, long-anticipated nova outburst should occur during the 2024-2026 period. Here, we present results of high-resolution optical spectroscopy of T~CrB in the period 2016 -- 2023. In these spectra, we measured the equivalent widths of the H$\alpha$, H$\beta$, HeI and HeII emission lines. The maximum equivalent width (EW) was recorded on May 2021, when the EW of $H\alpha$ reached  -44.6~\AA\ and H$\beta$ = -21.5~\AA. At the other extreme, the minimum of EW($H\alpha$)= -2.9~\AA\ was recorded in October 2023. After October 2023, the B-band emission brightened, suggesting a
re-appearance of the orbital modulation.
In addition to the optical data, we study the X-ray behaviour in the same period. We find a strong correlation between $EW(H\alpha)$ and X-ray flux with a correlation coefficient -0.78 and a significance of 2.6$\times 10^{-5}$. }

\keywords{stars: novae, cataclysmic variables -- binaries: symbiotic -- stars: individual: T~CrB}

%

\section{Introduction}
T~CrB is a symbiotic binary system that consists of an M4 giant star (M{\"u}rset \& Schmid 1999) that fills its Roche lobe (Belczy{\'n}ski \& Miko{\l}ajewska 1998) and a massive
white dwarf (M$_{WD}$ = 1.37 $\pm$ 0.13~$M_\odot$; Stanishev et al. 2004). Its orbital period is 227.57~d (Fekel et al. 2000; Schaefer 2023a). T~CrB is a recurrent nova with recorded outbursts in 1217, 1787, 1866 and 1946 with a recurrence time between the outbursts of roughly 80~yr (Sanford 1949; Schaefer 2023b). The next nova outburst of T~CrB is expected to occur in the period 2025$\pm$1 (Luna et al. 2020; Schaefer 2023a; Schaefer et al. 2023). The distance to T~CrB is 914~pc (Schaefer 2022).


In 2015, T~CrB  entered into a super-active stage; an stage similar to the one observed a few years before its nova eruption in 1946 (Munari et al. 2016; I{\l}kiewicz et al. 2016; Luna et al. 2018).
This stage is characterized by a brightening in UV/optical and radio wavelengths, disappearance of the orbital modulation in the B-band light curve, appearance of strong and high ionization emission lines, while the hard X-ray emission almost completely vanished (Munari et al. 2016, Luna et al. 2018; Linford et al. 2019). The super-active stage of T~CrB probably is connected with a SU~UMa dwarf nova-type activity (I{\l}kiewicz, Miko{\l}ajewska \& Stoyanov 2023). 

Observations since mid-2023 of T~CrB point toward the end of the super-active stage (Munari 2023; Teyssier et al. 2023) and the system is now in the pre-eruption dip state (Schaefer et al. 2023; Kuin et al. 2023).
Here we present an analysis of the high-resolution optical spectroscopy of T~CrB taken in the period 2016 -- 2023, which covers the super-active stage. 


\section{Observations}

In the period 2016--2023, we obtained 28 spectra with the ESpeRo fiber-fed Echelle spectrograph (Bonev et al. 2017) mounted on the 2m telescope of the Rozhen National Astronomical Observatory, Bulgaria. The ESpeRo spectrograph has a resolution of $\sim$ 30~000 and covers the range 3900~\AA\ -- 9000~\AA. The spectra were reduced following standard procedures, including bias removal, flat-field correction, wavelength calibration and correction for the Earth's motion. 
The spectra obtained within each observational night are processed and measured independently. On each spectrum we measured the equivalent widths of the H$\alpha$, H$\beta$, HeI~6678~\AA\ and HeII~4686~\AA\ emission lines. The measured EWs are listed on Table~\ref{obs}.

Since its launch, the {\it Neil Gehrels Swift Observatory} has been performing a set of Target of Opportunities observations of T~CrB in order to follow the X-ray evolution. Some of these observations have been already analyzed and presented in Luna et al. (2018) and Kuin et al. (2023). We used the long-term X-ray light curve to study how it correlates with the behavior observed in the optical spectra presented here. The X-ray light curve was constructed using the web tool \url{https://www.swift.ac.uk/user_objects/} (Evans et al. 2009). 

\begin{table}[]
    \centering
    \begin{tabular}{|c|c|c|c|c|c|c}
    \hline
date       & MJD                 & EW(H$\alpha$) &  EW(H$\beta$) & EW(HeI~6678) & EW(HeII~4686)  & \\
yyyy-mm-dd &                     & [\AA]         &  [\AA]  &  [\AA] & [\AA] & \\
      \hline
2016-03-30 &	57478.49002312   &     -35.1 &   -22.9  &  -8.3  &   -9.1 & \\
2017-03-17 &	57830.50380812   &     -39.7 &   -26.6  &  -5.3  &   -5.4 & \\
2017-07-02 &	57937.33295112   &     -41.5 &   -23.9  &  -6.9  &   -7.3 & \\
2017-12-30 &	58118.65019712   &     -39.4 &   -22.9  &  -6.5  &   -7.6 & \\
2018-04-02 &	58211.45186312   &     -43.5 &   -27.1  &  -6.7  &   -9.5 & \\
2018-08-30 &	58361.29115712   &     -42.9 &   -22.7  &  -6.9  &   -8.1 & \\
2018-08-31 &	58362.27397012   &     -42.0 &   -23.6  &  -6.5  &   -8.3 & \\
2018-09-01 &	58363.27300912   &     -41.9 &   -23.3  &  -6.4  &   -8.5 & \\
2018-12-27 &	58480.61445612   &     -36.5 &   -18.8  &  -4.8  &   -7.0 & \\
2019-03-17 &	58560.50748812   &     -38.6 &   -20.7  &  -5.8  &   -8.6 & \\
2021-01-01 &	59216.65751212   &     -33.5 &   -14.0  &  -3.0  &   -4.1 & \\
2021-04-01 &	59306.38541670	 &     -38.4 &   -17.8  &  -5.5  &   -5.0 & \\
2021-05-23 &	59358.41083312   &     -44.6 &   -21.5  &  -5.1  &   -5.4 & \\
2021-06-21 &	59387.47768512   &     -27.1 &   -14.2  &  -2.8  &   -2.4 & \\
2021-09-17 &	59475.31023112   &     -32.9 &   -17.5  &  -5.0  &   -4.4 & \\
2022-02-10 &	59621.51474512   &     -36.6 &   -18.5  &  -4.9  &   -4.5 & \\
2022-03-12 &	59651.54438712   &     -30.2 &   -13.6  &  -3.2  &   -3.5 & \\
2022-04-13 &	59683.38937512   &     -33.9 &   -17.6  &  -4.4  &   -6.9 & \\
2022-05-21 &	59721.34887712   &     -36.6 &   -20.8  &  -5.2  &   -4.7 & \\
2022-08-08 &	59800.40487312   &     -39.6 &   -20.7  &  -5.5  &   -5.0 & \\
2022-09-14 &	59837.25776612   &     -39.7 &   -20.1  &  -4.1  &   -4.5 & \\
2023-01-03 &	59948.64393512   &     -29.6 &   -18.6  &  -3.6  &   -3.7 & \\
2023-02-09 &	59985.56171312   &     -29.6 &   -16.6  &  -3.9  &   -3.6 & \\
2023-03-29 &	60033.43622712   &     -32.9 &   -20.7  &  -3.5  &   -4.6 & \\
2023-04-12 &	60047.39773112   &     -24.7 &   -13.5  &  -1.7  &   -1.9 & \\
2023-06-07 &	60103.37072912   &     -10.1 &   -6.7   &  --   &   --  & \\
2023-10-23 &	60241.19274000   &     -2.9  &  --     &  --   &   --  & \\
2023-12-25 &	60304.64093286   &     -7.3  &  -1.4    &  --   &   --  & \\
\hline
    \end{tabular}
    \caption{Log of the optical observations of T~CrB. In the table are given the date of the observation, the MJD of the observation and the measured EWs of the H$\alpha$, H$\beta$, HeI~6678~\AA\ and HeII~4686~\AA\ emission lines.}
    \label{obs}
\end{table}

\section{Results}

   \begin{figure*}
   \centering
   \includegraphics[width=6cm]{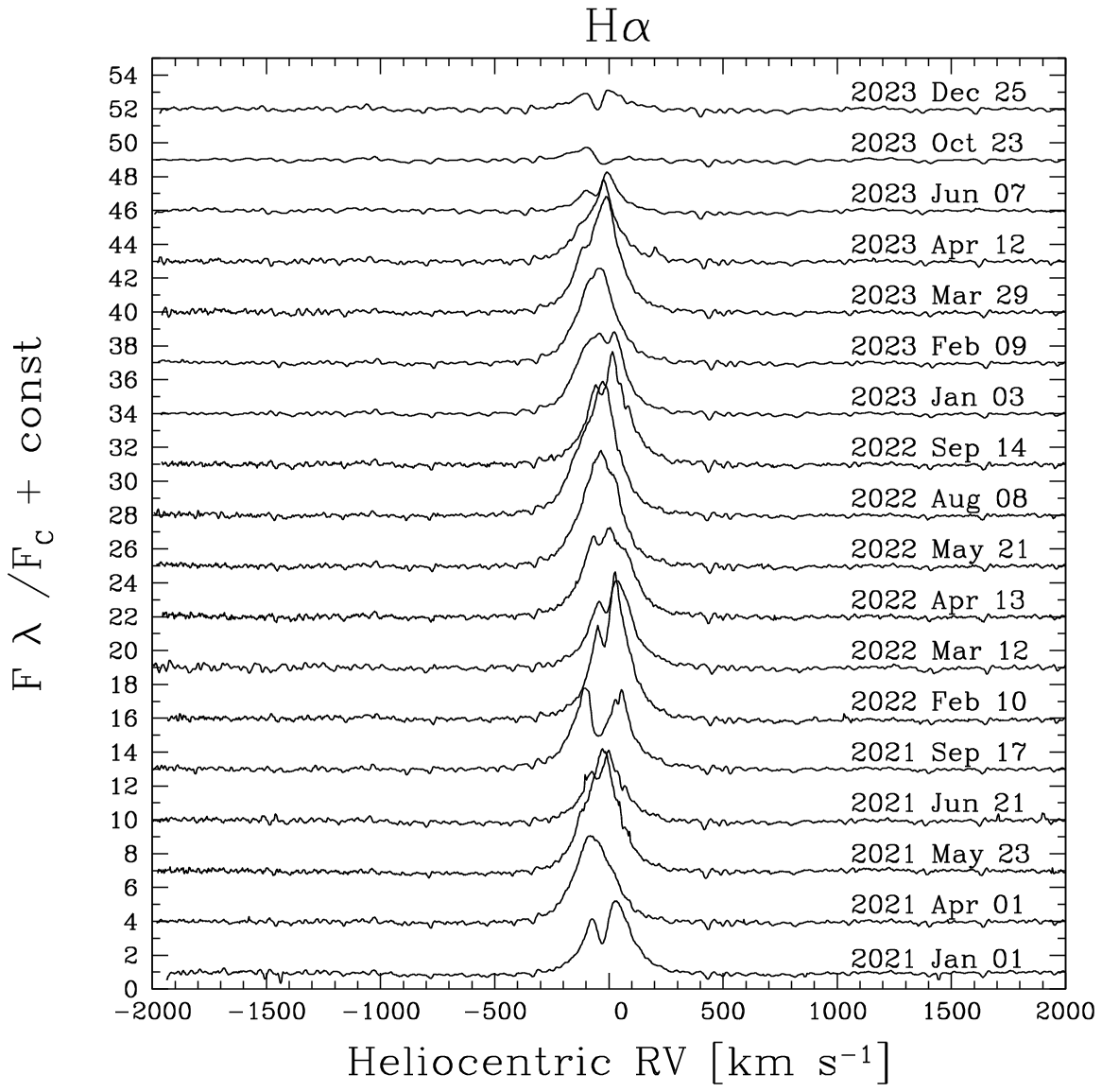}
   \includegraphics[width=6cm]{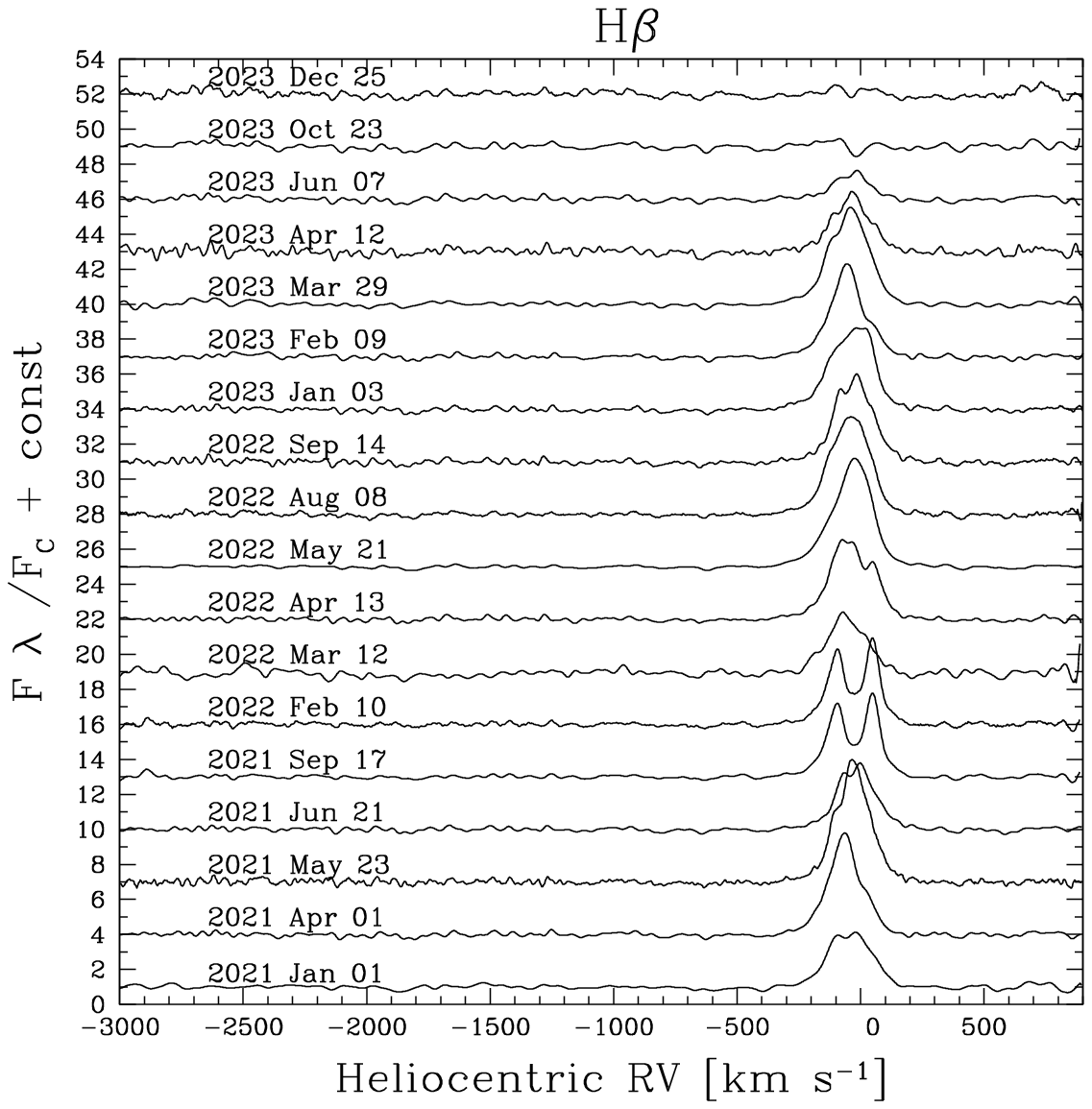}
   \includegraphics[width=10cm]{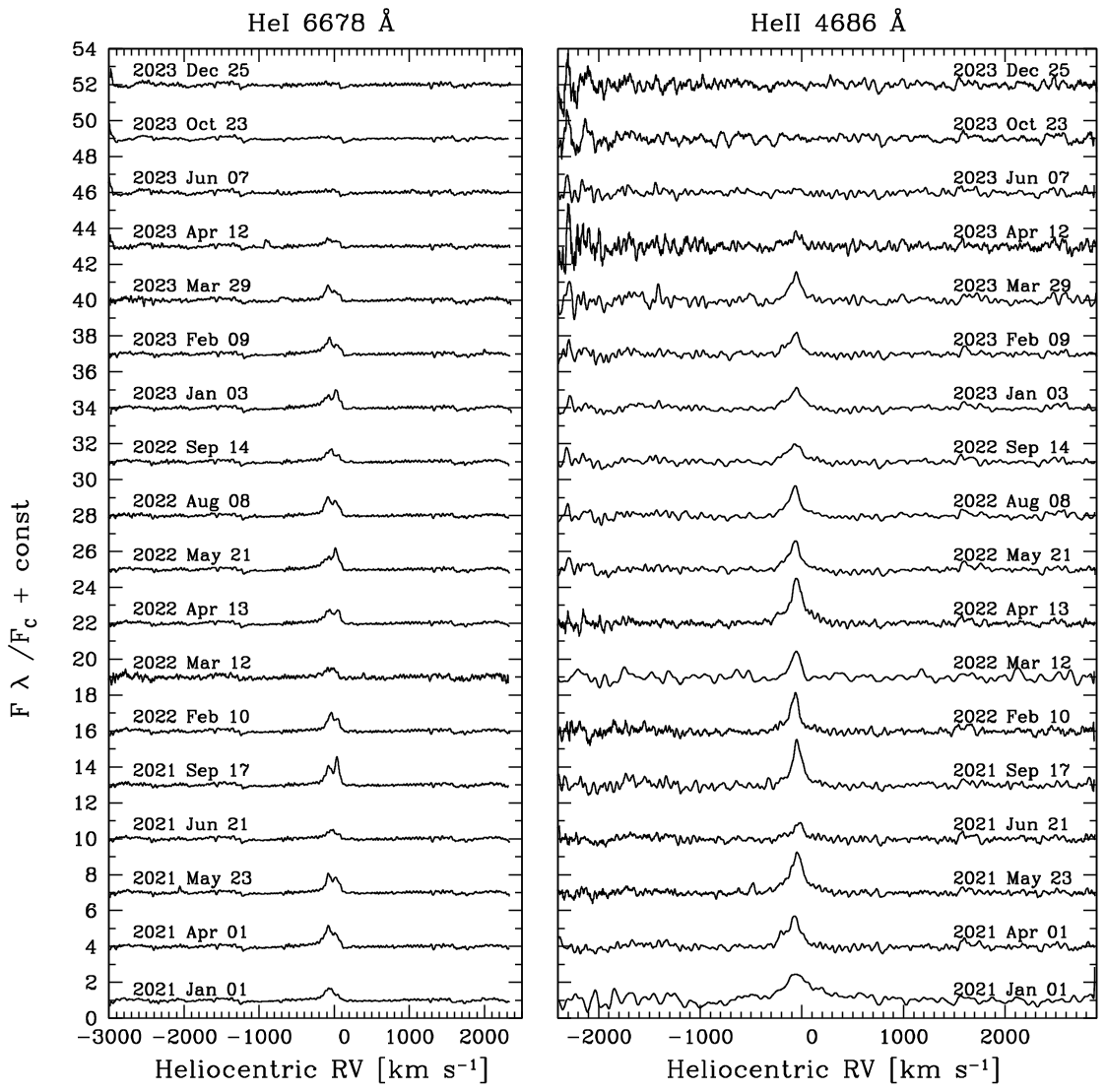}
   \caption{On the upper panel are present the H$\alpha$ and the H$\beta$ emission lines. On the bottom panel are the HeI~6678~\AA\ and HeII~4686~\AA\ emission lines.}
              \label{spec}%
    \end{figure*}

In Figure \ref{spec} we present the evolution of the emission lines that we measured in the spectra of T~CrB. In the period of the observations, H$\alpha$ and H$\beta$ are highly variable going from a single-peaked to double-peaked structure and even a triple-peaked profile of the H$\alpha$ line in 2023 April. The EW of H$\alpha$ line varies from -45~\AA\ to -3~\AA. The EW of the H$\beta$ emission line varies in the range -21~\AA\ -- -1.4~\AA. The He emission lines are also strongly variable. The EW of the HeI~6678~\AA\ line varies from -5.5~\AA\ to -2~\AA\ and the EW of the HeII~4686~\AA\ -- from -7~\AA\ to -2~\AA. The He lines are not detected in the last spectra taken since June 2023. \\

   \begin{figure}
   \centering
   \includegraphics[width=10cm]{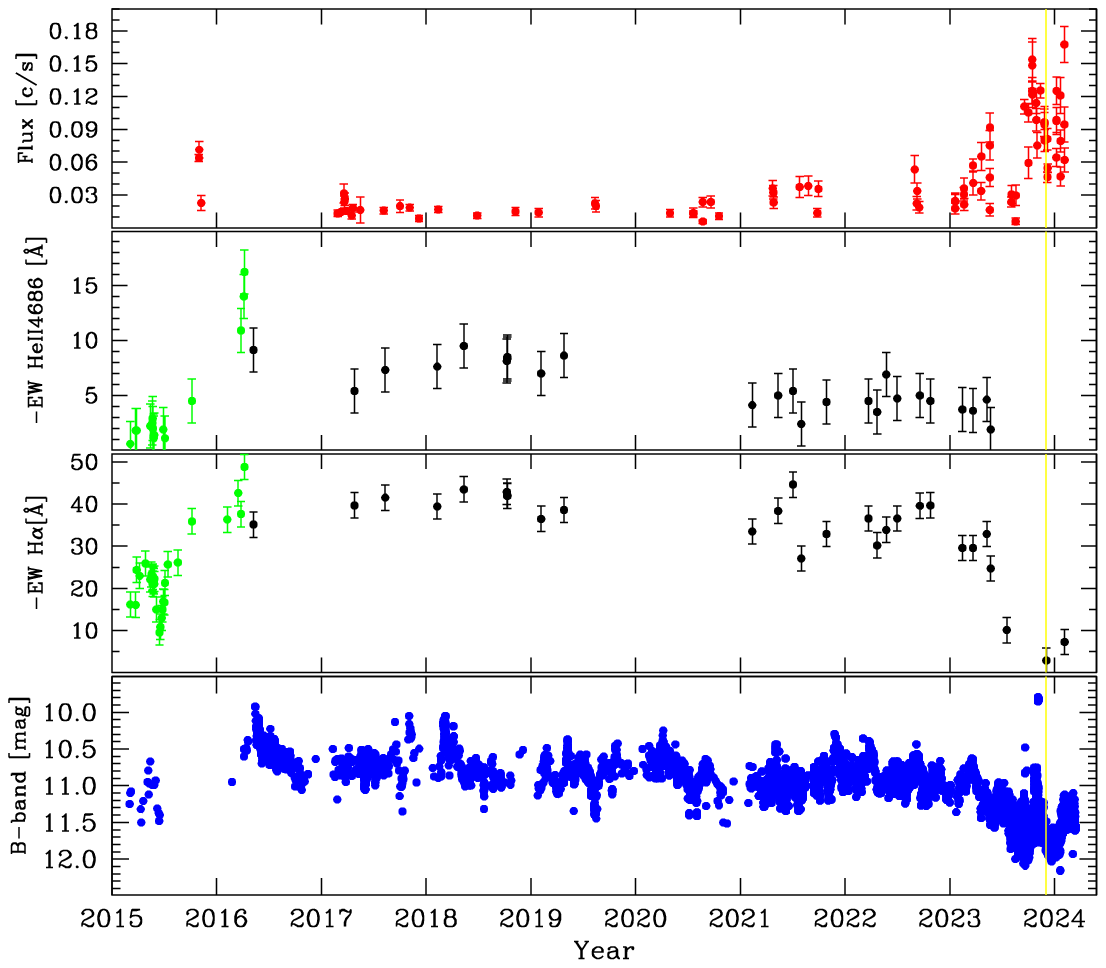}
      \caption{Long-term variability of T~CrB. From top to bottom, the X-ray activity, the EWs of HeII~4686$\AA$ and H$\alpha$, respectively, and the B-band light curve from AAVSO data. The green symbols represent the data taken from I{\l}kiewicz et al.(2016). The yellow line marks the date when we detected the minimum EW of the H$\alpha$ line.}
         \label{variability}
   \end{figure}

In Figure \ref{variability} we present the long-term variability of T~CrB from 2015 to 2023. In the figure we plot the X-ray light curve in the 0.3-10 keV energy range, the EWs of HeII~4686$\AA$ and H$\alpha$ respectively, and the B-band light curve based on AAVSO data. For the EW data, in addition to the Rozhen spectral observations, we also used data published in I{\l}kiewicz et al.(2016). Regardless of the orbital variability, the long-term behavior shows a steady decrease in the B-band magnitude and the EWs of the emission lines since the maximum in March 2016. As the B-band brightness is dominated by an emission from the accretion disc, the decrease of the brightness suggests that the accretion rate has decreased to pre-super active stage levels. During the same period, the amplitude of the optical flickering from T~CrB also increased (Shore et al. 2023; Minev et al. 2023). 
The EWs of the Balmer lines reach a minimum in 2023 October, and a slight increase of the EWs is present in 2023 December. In addition to the increase of the EWs, there is an increase in the B-band brightness due to the orbital modulation.

   \begin{figure}
   \centering
   \includegraphics[width=10cm]{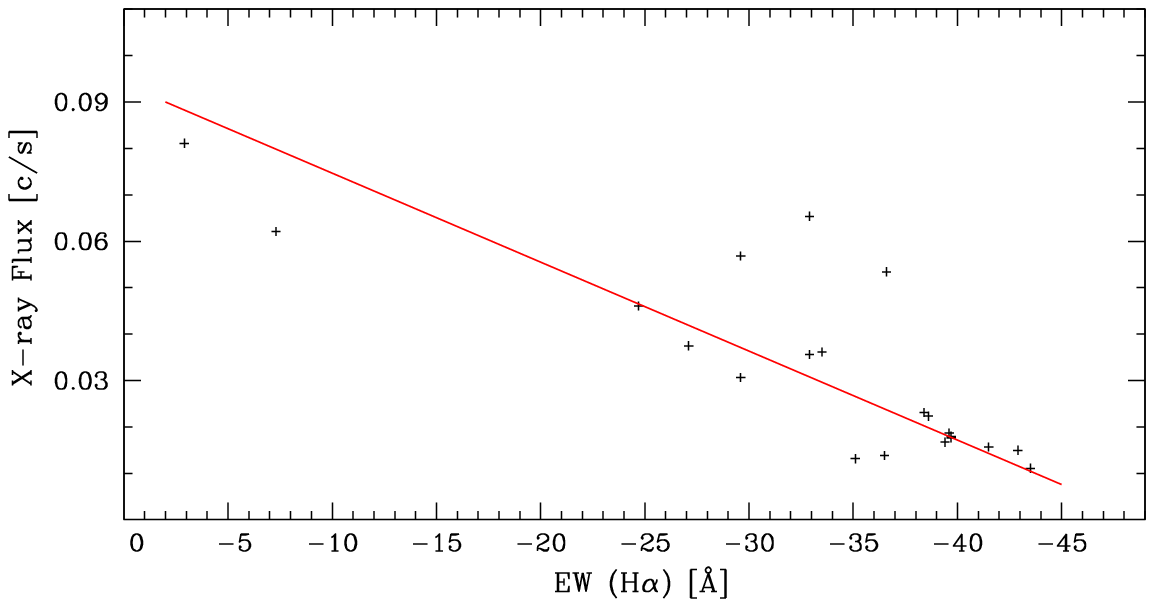}
   \vspace*{-5cm}
      \caption{The equivalent widths (EWs) of the H$\alpha$ versus the X-ray flux. The solid red line is the best linear fit.}
         \label{correlation}
   \end{figure}

In Fig.~\ref{correlation} we plot EW(H$\alpha$) versus the X-ray flux of T~CrB. The linear fit (of type $y = a + bx$) gives: 

\begin{equation}
   {\rm X-ray~Flux} = 0.0939 [\pm 0.0008] - 0.00192 [\pm 0.00002] EW(H\alpha).
\end{equation}

The errors of the coefficients are given in brackets. The Pearson correlation coefficient is 0.81.
The Spearman ($\rho$) rank correlation gives $\rho$ = -0.78 (significance 2.6 $\times$ 10$^{-5}$).
The analysis is indicating that the correlation between the EW(H$\alpha$) and the X-ray flux 
is strong and highly significant.

\section{Discussion} 
The EWs of the emission lines reached a maximum in March 2016, when EW($H\alpha$)= -48~\AA\ 
(I{\l}kiewicz et al., 2016). Since then, there follows a slow decrease. 
In our dataset, the maximum of the EW($H\alpha$) = -44.6~\AA\ is in May 2021.
The maximum values of the EW($H\alpha$) were in the range $-40 \div -44$~\AA\ during the period 
March 2017 -- September 2022, when the mass accretion rate was $2 \div 3 \times 10^{-8}
~M_\odot~yr^{-1}$ (Zamanov et al. 2023). In this period, the hard X-ray flux was low -- $\sim$
0.02~c/s.
Meanwhile, the minimum of the H$\alpha$ emission is in October 2023, when the EW is only
-2.9~\AA. The mass accretion rate in this period is less than $0.3 \times 10^{-8}
~M_\odot~yr^{-1}$, which means that the mass accretion rate is at least ten times lower than the
rate during the period of activity.

By the end of the super-active stage, the X-ray flux has increased steadily. This suggests that the accretion rate decreased and that the boundary layer has now become mostly optically thin. With a lower mass accretion rate, the whole disc is becoming cooler, and thus red (Zamanov et al. 2023).

In the symbiotic stars, portion of the stellar wind of the red giant is photoionized by the hot companion resulting an ionized nebula giving a rise of bright optical and UV spectral lines (Kenyon \& Webbink 1984). The X-ray, UV and radio observations of T~CrB during the active stage reveal an increase of the ionizing radiation from the hot component (Luna et al. 2018; Luna et al. 2019; Linford et al. 2019). The ionization of a large portion of the nebula can cause an increase in the EW of the emission lines on the optical spectrum of T~CrB. The B-band emission is dominated by the emission from the accretion disc and an irradiation of the red giant by the white dwarf. These means that the decrease of the mass accretion rate leads to an decrease of the energy generated in the accretion disc which in not enough to ionize the nebula around the system.  

\section*{Conclusions} 
We obtained 28 high-resolution spectra of the symbiotic recurrent nova T~CrB in the period 
2016-2023. We measured the EWs of $H\alpha$, $H\beta$, HeI~6678 and HeII~4686 emission lines in
the spectra. The maximum of the emission lines is in May 2021 when the $H\alpha$ emission line is
reaching EW of -44.6~\AA\ and the EW($H\beta$)= -21.5~\AA. The minimum of EW($H\alpha$)= -2.9~\AA\
is in October 2023. After the end of the super-active stage of T~CrB, the hard X-rays are 
increasing, suggesting that the boundary layer is becoming optically thin. Moreover, the orbital 
modulation in B-band that was not present during the active stage, re-appeared again in the end of 
2023. We find a strong correlation between $EW(H\alpha)$ and X-ray flux with a correlation 
coefficient -0.78 and significance 2.6$\times 10^{-5}$. \\

{{\bf Acknowledgments:}
   We are grateful to an anonymous referee for the helpful comments and suggestions. GJML is member of the CIC-CONICET (Argentina) and acknowledges the support of the ANPCYT-PICT 0901/2017 grant. KI acknowledges support from Polish National Science Center grant 2021/40/C/ST9/00186.  YMN acknowledges support by the Bulgarian National Science Fund – project K$\Pi$-06-H58/3. We acknowledge with thanks the variable star observations from the AAVSO International Database contributed by observers worldwide and used in this research. The authors acknowledge the use of public data from the $Swift$ data archive, as well as the frequent ToO observations performed upon our request. Part of the research infrastructure this research was done with is funded by the Ministry of Education and Science of Bulgaria (support for the Bulgarian National Roadmap for Research Infrastructure).  }

%

\begin{thebibliography}{}
\bibitem[Belczynski \& Mikolajewska(1998)]{1998MNRAS.296...77B} Belczynski, K. \& Mikolajewska, J.\ 1998, \mnras, 296, 77
\bibitem[Bonev et al.(2017)]{2017BlgAJ..26...67B} Bonev, T., Markov, H., Tomov, T., et al.\ 2017, Bulgarian Astronomical Journal, 26, 67
\bibitem[Evans et al.(2009)]{2009MNRAS.397.1177E} Evans, P.~A., Beardmore, A.~P., Page, K.~L., et al.\ 2009, \mnras, 397, 1177
\bibitem[Fekel et al.(2000)]{2000AJ....119.1375F} Fekel, F.~C., Joyce, R.~R., Hinkle, K.~H., et al.\ 2000, \aj, 119, 1375
\bibitem[I{\l}kiewicz et al.(2023)]{2023ApJ...953L...7I} I{\l}kiewicz, K., Miko{\l}ajewska, J., \& Stoyanov, K.~A.\ 2023, \apjl, 953, L7
\bibitem[I{\l}kiewicz et al.(2016)]{2016MNRAS.462.2695I} I{\l}kiewicz, K., Miko{\l}ajewska, J., Stoyanov, K., et al.\ 2016, \mnras, 462, 2695
\bibitem[Kenyon \& Webbink(1984)]{1984ApJ...279..252K} Kenyon, S.~J. \& Webbink, R.~F.\ 1984, \apj, 279, 252
\bibitem[Kuin et al.(2023)]{2023ATel16114....1K} Kuin, N.~P., Luna, G.~J.~M., Page, K., et al.\ 2023, The Astronomer's Telegram, 16114
\bibitem[Linford et al.(2019)]{2019ApJ...884....8L} Linford, J.~D., Chomiuk, L., Sokoloski, J.~L., et al.\ 2019, \apj, 884, 8
\bibitem[Luna et al.(2018)]{2018A&A...619A..61L} Luna, G.~J.~M., Mukai, K., Sokoloski, J.~L., et al.\ 2018, \aap, 619, A61
\bibitem[Luna et al.(2019)]{2019ApJ...880...94L} Luna, G.~J.~M., Nelson, T., Mukai, K., et al.\ 2019, \apj, 880, 94
\bibitem[Luna et al.(2020)]{2020ApJ...902L..14L} Luna, G.~J.~M., Sokoloski, J.~L., Mukai, K., et al.\ 2020, \apjl, 902, L14
\bibitem[Minev et al.(2023)]{2023ATel16023....1M} Minev, M., Zamanov, R., \& Stoyanov, K.\ 2023, The Astronomer's Telegram, 16023
\bibitem[Munari(2023)]{2023RNAAS...7..145M} Munari, U.\ 2023, Research Notes of the American Astronomical Society, 7, 145
\bibitem[Munari et al.(2016)]{2016NewA...47....7M} Munari, U., Dallaporta, S., \& Cherini, G.\ 2016, \na, 47, 7
\bibitem[M{\"u}rset \& Schmid(1999)]{1999A&AS..137..473M} M{\"u}rset, U. \& Schmid, H.~M.\ 1999, \aaps, 137, 473
\bibitem[Sanford(1949)]{1949ApJ...109...81S} Sanford R.~F., 1949, ApJ, 109, 81
\bibitem[Schaefer(2022)]{2022MNRAS.517.6150S} Schaefer, B.~E.\ 2022, \mnras, 517, 6150
\bibitem[Schaefer(2023a)]{2023MNRAS.524.3146S} Schaefer, B.~E.\ 2023a, \mnras, 524, 3146
\bibitem[Schaefer(2023b]{2023arXiv230813668S} Schaefer B.~E., 2023b, Journal for the History of Astronomy, 54, 436
\bibitem[Schaefer et al.(2023)]{2023ATel16107....1S} Schaefer, B.~E., Kloppenborg, B., Waagen, E.~O., et al.\ 2023, The Astronomer's Telegram, 16107
\bibitem[Shore et al.(2023)]{2023ATel15916....1S} Shore, S.~N., Teyssier, F., \& ARAS Group\ 2023, The Astronomer's Telegram, 15916
\bibitem[Stanishev et al.(2004)]{2004A&A...415..609S} Stanishev, V., Zamanov, R., Tomov, N., et al.\ 2004, \aap, 415, 609
\bibitem[Teyssier et al.(2023)]{2023ATel16109....1T} Teyssier, F., Hinnefeld, J.~D., Boussin, C., et al.\ 2023, The Astronomer's Telegram, 16109
\bibitem[Zamanov et al.(2023)]{2023A&A...680L..18Z} Zamanov, R., Boeva, S., Latev, G.~Y., et al.\ 2023, \aap, 680, L18
\end{thebibliography}
%

\end{document}